\documentclass[showpacs,aps,prd]{revtex4}
\input epsf

\begin{document}

\title{$0^{+}$ fully-charmed tetraquark states}
\author{Jian-Rong Zhang}
\affiliation{Department of Physics, College of Liberal Arts and Sciences, National University of Defense Technology,
Changsha 410073, Hunan, People's Republic of China}


\begin{abstract}
Motivated by the LHCb's new observation of structures
in the $J/\psi$-pair invariant mass spectrum, for which could be classified
as possible $cc\bar{c}\bar{c}$ tetraquark candidates, we systematically study
$0^{+}$ fully-charmed tetraquark states through QCD sum rules.
Making the development of
calculation techniques to fourfold heavy hadronic systems,
four different configuration currents with $0^{+}$
are considered and vacuum condensates up to dimension
$6$ are included in the operator product expansion (OPE).
Finally, mass values acquired
for $0^{+}$ $cc\bar{c}\bar{c}$ tetraquark states agree well with the experimental data
of the broad structure,
which supports that it could be a $0^{+}$ fully-charmed tetraquark state.
\end{abstract}
\maketitle

\section{Introduction}\label{sec1}
By far, the topic of fully-charmed tetraquark state has attracted much attention.
For example, a variety of phenomenological models were
employed to predict the existence
of some states merely made up of four heavy quarks \cite{Predict-1,Predict-2,Predict-3,Predict-4,Predict-5,Predict-6,Predict-7,Predict-8,Predict-9,Predict-10,
Predict-11,Predict-12,Predict-13,Predict-14,Predict-15,Predict-16,Predict-16-a,Predict-17,Predict-18,Predict-19,Predict-20,
Predict-21,Predict-22,Predict-23,Predict-24,Predict-25,Predict-26,Predict-27}.
Particularly, without any light quark
contamination, fully-charmed tetraquark states are
ideal prototypes to refine one's understanding on heavy quark dynamics.

Recently, the invariant mass spectrum of double-$J/\psi$ was researched
using proton-proton collision data recorded by the LHCb experiment,
which shows a broad structure just above twice the $J/\psi$ mass ranging
from $6.2$ to $6.8~\mbox{GeV}$ and a narrower
structure around $6.9~\mbox{GeV}$, referred to as $X(6900)$ \cite{LHCb}.
Soon after the LHCb's new results, various investigations were presented
to explain them via different
approaches \cite{Th-1,Th-2,Th-2-a,Th-2-aa,Th-3,Th-4,Th-5,Th-6,Th-7,Th-8,Th-9,Th-10,
Th-11,Th-12,Th-13,Th-14,Th-15,Th-16,Th-17,Th-18,Th-19,Th-20,Th-21,Th-22,Th-23,Th-24,Th-25,Th-26}.
To probe a real hadron, one inevitably has to
face the sophisticated nonperturbative QCD problem.
As one trustable way for evaluating nonperturbative effects,
the QCD sum rule \cite{svzsum} is firmly established on the basic theory,
and has been widely applied to hadronic systems (for reviews see Refs.
\cite{overview,overview1,overview2,overview3} and references
therein). In particular on $0^{+}$ $cc\bar{c}\bar{c}$ tetraquark states, there are some
existing works \cite{Predict-11,Predict-12,Th-2,Th-10,Th-19} in different versions of QCD sum rules.

By comparison, Ref. \cite{Predict-11} used a moment QCD sum rule method augmented by fundamental
inequalities to explore the doubly hidden-charm/bottom tetraquark
states; Working with the Finite Energy version of the
QCD Inverse Laplace sum rules, Ref. \cite{Th-10}
investigated doubly-hidden scalar heavy molecules and tetraquarks states;
Besides, there have
some other QCD sum rule analysis of fully-heavy
tetraquark states involving
condensate contributions up to dimension $4$ in the OPE,
specially choosing the axial vector-axial vector configuration current in Ref. \cite{Predict-12},
paying attention to the scalar's first radial excited states \cite{Th-2}, or
introducing a relative P-wave to the diquark
operator of the tetraquark current \cite{Th-19}.
Consumingly motivated by the so exciting and significant structures observed
in the di-$J/\psi$ mass spectrum, we would follow
the previous QCD sum rule studies \cite{Nielsen,Nielsen1,Zhang} on hadrons
containing one or two heavy quarks
and devote to developing the corresponding calculation techniques to
fourfold heavy hadronic systems.
And then, we intend to systematically study $0^{+}$ fully-charmed
tetraquark states with QCD sum rules,
by taking into account four possible configuration currents
and calculating condensates up to dimension $6$.

The paper is organized as follows. After the Introduction, the QCD sum rule
is derived for $0^{+}$ fully-charmed tetraquark states in Sec. \ref{sec2},
along with numerical analysis and discussions in Sec.
\ref{sec3}. The last part contains a brief summary.

\section{$0^{+}$ fully-charmed tetraquark state QCD sum rules}\label{sec2}
Considering a tetraquark state, its interpolating current can ordinarily be
represented by
a diquark-antidiquark configuration.
Thus, following forms of currents could be constructed for $0^{+}$ $cc\bar{c}\bar{c}$
tetraquark states, with
\begin{eqnarray}
j&=&(Q_{a}^{T}C\gamma_{5}Q_{b})(\bar{Q}_{a}\gamma_{5}C\bar{Q}_{b}^{T})\nonumber
\end{eqnarray}
for the scalar-scalar configuration,
\begin{eqnarray}
j&=&(Q_{a}^{T}CQ_{b})(\bar{Q}_{a}C\bar{Q}_{b}^{T})\nonumber
\end{eqnarray}
for the pseudoscalar-pseudoscalar one,
\begin{eqnarray}
j&=&(Q_{a}^{T}C\gamma_{\mu}Q_{b})(\bar{Q}_{a}\gamma^{\mu}C\bar{Q}_{b}^{T})\nonumber
\end{eqnarray}
for the axial vector-axial vector (shortened to axial-axial) one, and
\begin{eqnarray}
j&=&(Q_{a}^{T}C\gamma_{5}\gamma_{\mu}Q_{b})(\bar{Q}_{a}\gamma^{\mu}\gamma_{5}C\bar{Q}_{b}^{T})\nonumber
\end{eqnarray}
for the vector-vector one.
Here the index $T$ indicates matrix
transposition, $C$ means the charge conjugation matrix,
$Q$ is the heavy charm quark, as well as $a$ and $b$ are color indices.
It is needed to state that these overall scalar forms of currents
are constructed as existing works
mainly taking into consideration
that all of their Lorentz indices being contracted.
One should note that
the situation for axial-axial and vector-vector could be more complicated
while characterizing a scalar state.
In a general way, the combination of
two objects with spin $1$ may have total spin equal to $0$, $1$ and $2$,
and the currents for axial-axial and vector-vector may represent a mixture of spins.
To select the spin $0$ part of these two currents,
one could try to
project their correlation functions into the spin $0$ state
with the help of the appropriate projection operators,
for which have been introduced in Refs. \cite{remark,remark1}.

To derive QCD sum rules,
one can start with
the two-point correlator
\begin{eqnarray}
\Pi(q^{2})=i\int
d^{4}x\mbox{e}^{iq.x}\langle0|T[j(x)j^{\dag}(0)]|0\rangle.
\end{eqnarray}
In phenomenology, it can be expressed as
\begin{eqnarray}\label{ph}
\Pi(q^{2})=\frac{\lambda_{H}^{2}}{M_{H}^{2}-q^{2}}+\frac{1}{\pi}\int_{s_{0}}
^{\infty}\frac{\mbox{Im}\big[\Pi^{\mbox{phen}}(s)\big]}{s-q^{2}}ds,
\end{eqnarray}
in which $M_{H}$ is the hadron's mass, $s_0$ denotes the continuum threshold,
and $\lambda_{H}$ displays
the coupling of the current to the hadron $\langle0|j|H\rangle=\lambda_{H}$.
Moreover, it can theoretically be rewritten as
\begin{eqnarray}\label{ope}
\Pi(q^{2})=\int_{(4m_{Q})^{2}}^{\infty}\frac{\rho}{s-q^{2}}ds,
\end{eqnarray}
where $m_{Q}$ is the heavy charm mass,
and the spectral density $\rho=\frac{1}{\pi}\mbox{Im}\big[\Pi(s)\big]$.
After equating Eqs. (\ref{ph}) and (\ref{ope}), adopting quark-hadron duality, and
applying a Borel transform, it yields
\begin{eqnarray}\label{sumrule}
\lambda_{H}^{2}e^{-M_{H}^{2}/M^{2}}&=&\int_{(4m_{Q})^{2}}^{s_{0}}\rho e^{-s/M^{2}}ds,
\end{eqnarray}
with $M^2$ the Borel parameter.
Taking
the derivative of Eq. (\ref{sumrule}) with respect to $-\frac{1}{M^2}$ and then dividing
the result by Eq. (\ref{sumrule}) itself, one could achieve the mass sum rule
\begin{eqnarray}\label{sum rule}
M_{H}&=&\sqrt{\int_{(4m_{Q})^{2}}^{s_{0}}\rho s
e^{-s/M^{2}}ds/
\int_{(4m_{Q})^{2}}^{s_{0}}\rho e^{-s/M^{2}}ds}.
\end{eqnarray}

In the OPE calculation, one could work
at the momentum-space making use of the heavy-quark propagator
\cite{reinders}, and
then the result is dimensionally
regularized at $D=4$, by making an extension of the calculation techniques \cite{Nielsen,Nielsen1,Zhang}
to fully heavy tetraquark systems.
The spectral density is concretely expressed as
$\rho=\rho^{\mbox{pert}}+\rho^{\langle
g^{2}G^{2}\rangle}+\rho^{\langle
g^{3}G^{3}\rangle}$, with
\begin{eqnarray}
\rho^{\mbox{pert}}&=&-\frac{1}{2^{10}\pi^{6}}\int_{\alpha_{min}}^{\alpha_{max}}\frac{d\alpha}{\alpha^{3}}\int_{\beta_{min}}^{\beta_{max}}\frac{d\beta}{\beta^{3}}\int_{\gamma_{min}}^{\gamma_{max}}\frac{d\gamma}{\gamma^{3}}
\frac{1}{\textbf{H}^{6}}\bigg[m_{Q}^{2}-\frac{(1-\alpha-\beta-\gamma)s}{\textbf{H}}\bigg]^{2}\nonumber\\
&\times&
\bigg\{\Big[-3(1-\alpha-\beta-\gamma)+4\Big(\gamma(1-\alpha-\beta-\gamma)+\alpha\beta\Big)\textbf{H}
-6\alpha\beta\gamma\textbf{H}^{2}\Big]\textbf{H}^{2}m_{Q}^{4}\nonumber\\
&+&(1-\alpha-\beta-\gamma)\Big[18(1-\alpha-\beta-\gamma)-10\Big(\gamma(1-\alpha-\beta-\gamma)+\alpha\beta\Big)\textbf{H}\Big]\textbf{H}m_{Q}^{2}s
-21(1-\alpha-\beta-\gamma)^{3}s^{2}\bigg\},\nonumber
\end{eqnarray}

\begin{eqnarray}
\rho^{\langle g^{2}G^{2}\rangle}&=&-\frac{m_{Q}^{2}\langle g^{2}G^{2}\rangle}{3\cdot2^{9}\pi^{6}}\int_{\alpha_{min}}^{\alpha_{max}}\frac{d\alpha}{\alpha^{3}}\int_{\beta_{min}}^{\beta_{max}}\frac{d\beta}{\beta^{3}}\int_{\gamma_{min}}^{\gamma_{max}}\frac{d\gamma}{\gamma^{3}}
\frac{(1-\alpha-\beta-\gamma)^{2}}{\textbf{H}^{5}}\nonumber\\
&\times&\bigg\{\Big[-6(1-\alpha-\beta-\gamma)^{2}
+(1-\alpha-\beta-\gamma)\Big(6\gamma+2\gamma(1-\alpha-\beta-\gamma)+2\alpha\beta\Big)\textbf{H}
-3\alpha\beta\gamma\textbf{H}^{2}\Big]\textbf{H}m_{Q}^{2}\nonumber\\
&+&3(1-\alpha-\beta-\gamma)^{2}\Big[4(1-\alpha-\beta-\gamma)-3\gamma\textbf{H}\Big]s\bigg\},\nonumber
\end{eqnarray}

\begin{eqnarray}
\rho^{\langle g^{3}G^{3}\rangle}&=&-\frac{\langle g^{3}G^{3}\rangle}{3\cdot2^{10}\pi^{6}}\int_{\alpha_{min}}^{\alpha_{max}}\frac{d\alpha}{\alpha^{3}}\int_{\beta_{min}}^{\beta_{max}}\frac{d\beta}{\beta^{3}}\int_{\gamma_{min}}^{\gamma_{max}}\frac{d\gamma}{\gamma^{3}}
\frac{(1-\alpha-\beta-\gamma)^{3}}{\textbf{H}^{5}}
\bigg\{\Big[-3(1-\alpha-\beta-\gamma)\nonumber\\
&-&6(1-\alpha-\beta-\gamma)^{2}+6\gamma(1-\alpha-\beta-\gamma)\textbf{H}
+\alpha\beta\textbf{H}\Big]\textbf{H}m_{Q}^{2}
+6(1-\alpha-\beta-\gamma)^{2}s\bigg\},\nonumber
\end{eqnarray}

for the scalar-scalar current,

\begin{eqnarray}
\rho^{\mbox{pert}}&=&-\frac{1}{2^{10}\pi^{6}}\int_{\alpha_{min}}^{\alpha_{max}}\frac{d\alpha}{\alpha^{3}}\int_{\beta_{min}}^{\beta_{max}}\frac{d\beta}{\beta^{3}}\int_{\gamma_{min}}^{\gamma_{max}}\frac{d\gamma}{\gamma^{3}}
\frac{1}{\textbf{H}^{6}}\bigg[m_{Q}^{2}-\frac{(1-\alpha-\beta-\gamma)s}{\textbf{H}}\bigg]^{2}\nonumber\\
&\times&
\bigg\{\Big[-3(1-\alpha-\beta-\gamma)-4\Big(\gamma(1-\alpha-\beta-\gamma)+\alpha\beta\Big)\textbf{H}
-6\alpha\beta\gamma\textbf{H}^{2}\Big]\textbf{H}^{2}m_{Q}^{4}\nonumber\\
&+&(1-\alpha-\beta-\gamma)\Big[18(1-\alpha-\beta-\gamma)
+10\Big(\gamma(1-\alpha-\beta-\gamma)+\alpha\beta\Big)\textbf{H}\Big]\textbf{H}m_{Q}^{2}s
-21(1-\alpha-\beta-\gamma)^{3}s^{2}\bigg\},\nonumber
\end{eqnarray}

\begin{eqnarray}
\rho^{\langle g^{2}G^{2}\rangle}&=&-\frac{m_{Q}^{2}\langle g^{2}G^{2}\rangle}{3\cdot2^{9}\pi^{6}}\int_{\alpha_{min}}^{\alpha_{max}}\frac{d\alpha}{\alpha^{3}}\int_{\beta_{min}}^{\beta_{max}}\frac{d\beta}{\beta^{3}}\int_{\gamma_{min}}^{\gamma_{max}}\frac{d\gamma}{\gamma^{3}}
\frac{(1-\alpha-\beta-\gamma)^{2}}{\textbf{H}^{5}}\nonumber\\
&\times&\bigg\{\Big[-6(1-\alpha-\beta-\gamma)^{2}
-(1-\alpha-\beta-\gamma)\Big(6\gamma+2\gamma(1-\alpha-\beta-\gamma)+2\alpha\beta\Big)\textbf{H}
-3\alpha\beta\gamma\textbf{H}^{2}\Big]\textbf{H}m_{Q}^{2}\nonumber\\
&+&3(1-\alpha-\beta-\gamma)^{2}\Big[4(1-\alpha-\beta-\gamma)+3\gamma\textbf{H}\Big]s\bigg\},\nonumber
\end{eqnarray}

\begin{eqnarray}
\rho^{\langle g^{3}G^{3}\rangle}&=&-\frac{\langle g^{3}G^{3}\rangle}{3\cdot2^{10}\pi^{6}}\int_{\alpha_{min}}^{\alpha_{max}}\frac{d\alpha}{\alpha^{3}}\int_{\beta_{min}}^{\beta_{max}}\frac{d\beta}{\beta^{3}}\int_{\gamma_{min}}^{\gamma_{max}}\frac{d\gamma}{\gamma^{3}}
\frac{(1-\alpha-\beta-\gamma)^{3}}{\textbf{H}^{5}}
\bigg\{\Big[-3(1-\alpha-\beta-\gamma)\nonumber\\
&-&6(1-\alpha-\beta-\gamma)^{2}-6\gamma(1-\alpha-\beta-\gamma)\textbf{H}
-\alpha\beta\textbf{H}\Big]\textbf{H}m_{Q}^{2}
+6(1-\alpha-\beta-\gamma)^{2}s\bigg\},\nonumber
\end{eqnarray}

for the pseudoscalar-pseudoscalar current,

\begin{eqnarray}
\rho^{\mbox{pert}}&=&-\frac{1}{2^{8}\pi^{6}}\int_{\alpha_{min}}^{\alpha_{max}}\frac{d\alpha}{\alpha^{3}}\int_{\beta_{min}}^{\beta_{max}}\frac{d\beta}{\beta^{3}}\int_{\gamma_{min}}^{\gamma_{max}}\frac{d\gamma}{\gamma^{3}}
\frac{1}{\textbf{H}^{6}}\bigg[m_{Q}^{2}-\frac{(1-\alpha-\beta-\gamma)s}{\textbf{H}}\bigg]^{2}\nonumber\\
&\times&
\bigg\{\Big[-3(1-\alpha-\beta-\gamma)+2\Big(\gamma(1-\alpha-\beta-\gamma)+\alpha\beta\Big)\textbf{H}
-6\alpha\beta\gamma\textbf{H}^{2}\Big]\textbf{H}^{2}m_{Q}^{4}\nonumber\\
&+&(1-\alpha-\beta-\gamma)\Big[18(1-\alpha-\beta-\gamma)-5\Big(\gamma(1-\alpha-\beta-\gamma)+\alpha\beta\Big)\textbf{H}\Big]\textbf{H}m_{Q}^{2}s
-21(1-\alpha-\beta-\gamma)^{3}s^{2}\bigg\},\nonumber
\end{eqnarray}

\begin{eqnarray}
\rho^{\langle g^{2}G^{2}\rangle}&=&-\frac{m_{Q}^{2}\langle g^{2}G^{2}\rangle}{3\cdot2^{8}\pi^{6}}\int_{\alpha_{min}}^{\alpha_{max}}\frac{d\alpha}{\alpha^{3}}\int_{\beta_{min}}^{\beta_{max}}\frac{d\beta}{\beta^{3}}\int_{\gamma_{min}}^{\gamma_{max}}\frac{d\gamma}{\gamma^{3}}
\frac{(1-\alpha-\beta-\gamma)^{2}}{\textbf{H}^{5}}\nonumber\\
&\times&
\bigg\{\Big[-12(1-\alpha-\beta-\gamma)^{2}
+(1-\alpha-\beta-\gamma)\Big(6\gamma+2\gamma(1-\alpha-\beta-\gamma)+2\alpha\beta\Big)\textbf{H}
-6\alpha\beta\gamma\textbf{H}^{2}\Big]\textbf{H}m_{Q}^{2}\nonumber\\
&+&3(1-\alpha-\beta-\gamma)^{2}\Big[8(1-\alpha-\beta-\gamma)-3\gamma\textbf{H}\Big]s\bigg\},\nonumber
\end{eqnarray}

\begin{eqnarray}
\rho^{\langle g^{3}G^{3}\rangle}&=&-\frac{\langle g^{3}G^{3}\rangle}{3\cdot2^{9}\pi^{6}}\int_{\alpha_{min}}^{\alpha_{max}}\frac{d\alpha}{\alpha^{3}}\int_{\beta_{min}}^{\beta_{max}}\frac{d\beta}{\beta^{3}}\int_{\gamma_{min}}^{\gamma_{max}}\frac{d\gamma}{\gamma^{3}}
\frac{(1-\alpha-\beta-\gamma)^{3}}{\textbf{H}^{5}}
\bigg\{\Big[-6(1-\alpha-\beta-\gamma)\nonumber\\
&-&12(1-\alpha-\beta-\gamma)^{2}+6\gamma(1-\alpha-\beta-\gamma)\textbf{H}
+\alpha\beta\textbf{H}\Big]\textbf{H}m_{Q}^{2}
+12(1-\alpha-\beta-\gamma)^{2}s\bigg\},\nonumber
\end{eqnarray}

for the axial-axial current, and

\begin{eqnarray}
\rho^{\mbox{pert}}&=&-\frac{1}{2^{8}\pi^{6}}\int_{\alpha_{min}}^{\alpha_{max}}\frac{d\alpha}{\alpha^{3}}\int_{\beta_{min}}^{\beta_{max}}\frac{d\beta}{\beta^{3}}\int_{\gamma_{min}}^{\gamma_{max}}\frac{d\gamma}{\gamma^{3}}
\frac{1}{\textbf{H}^{6}}\bigg[m_{Q}^{2}-\frac{(1-\alpha-\beta-\gamma)s}{\textbf{H}}\bigg]^{2}\nonumber\\
&\times&
\bigg\{\Big[-3(1-\alpha-\beta-\gamma)-2\Big(\gamma(1-\alpha-\beta-\gamma)+\alpha\beta\Big)\textbf{H}
-6\alpha\beta\gamma\textbf{H}^{2}\Big]\textbf{H}^{2}m_{Q}^{4}\nonumber\\
&+&(1-\alpha-\beta-\gamma)\Big[18(1-\alpha-\beta-\gamma)+5\Big(\gamma(1-\alpha-\beta-\gamma)+\alpha\beta\Big)\textbf{H}\Big]\textbf{H}m_{Q}^{2}s
-21(1-\alpha-\beta-\gamma)^{3}s^{2}\bigg\},\nonumber
\end{eqnarray}

\begin{eqnarray}
\rho^{\langle g^{2}G^{2}\rangle}&=&-\frac{m_{Q}^{2}\langle g^{2}G^{2}\rangle}{3\cdot2^{8}\pi^{6}}\int_{\alpha_{min}}^{\alpha_{max}}\frac{d\alpha}{\alpha^{3}}\int_{\beta_{min}}^{\beta_{max}}\frac{d\beta}{\beta^{3}}\int_{\gamma_{min}}^{\gamma_{max}}\frac{d\gamma}{\gamma^{3}}
\frac{(1-\alpha-\beta-\gamma)^{2}}{\textbf{H}^{5}}\nonumber\\
&\times&\bigg\{\Big[-12(1-\alpha-\beta-\gamma)^{2}
-(1-\alpha-\beta-\gamma)\Big(6\gamma+2\gamma(1-\alpha-\beta-\gamma)+2\alpha\beta\Big)\textbf{H}
-6\alpha\beta\gamma\textbf{H}^{2}\Big]\textbf{H}m_{Q}^{2}\nonumber\\
&+&3(1-\alpha-\beta-\gamma)^{2}\Big[8(1-\alpha-\beta-\gamma)+3\gamma\textbf{H}\Big]s\bigg\},\nonumber
\end{eqnarray}

\begin{eqnarray}
\rho^{\langle g^{3}G^{3}\rangle}&=&-\frac{\langle g^{3}G^{3}\rangle}{3\cdot2^{9}\pi^{6}}\int_{\alpha_{min}}^{\alpha_{max}}\frac{d\alpha}{\alpha^{3}}\int_{\beta_{min}}^{\beta_{max}}\frac{d\beta}{\beta^{3}}\int_{\gamma_{min}}^{\gamma_{max}}\frac{d\gamma}{\gamma^{3}}
\frac{(1-\alpha-\beta-\gamma)^{3}}{\textbf{H}^{5}}
\bigg\{\Big[-6(1-\alpha-\beta-\gamma)\nonumber\\
&-&12(1-\alpha-\beta-\gamma)^{2}-6\gamma(1-\alpha-\beta-\gamma)\textbf{H}
-\alpha\beta\textbf{H}\Big]\textbf{H}m_{Q}^{2}
+12(1-\alpha-\beta-\gamma)^{2}s\bigg\},\nonumber
\end{eqnarray}

for the vector-vector current. It is
defined as
\begin{eqnarray}
\textbf{H}&=&1+(1-\alpha-\beta-\gamma)\Big(\frac{1}{\alpha}+\frac{1}{\beta}+\frac{1}{\gamma}\Big),\nonumber\\
\alpha_{min}&=&\frac{1}{2}\Bigg[\bigg(1-\frac{8m_{Q}^{2}}{s}\bigg)-\sqrt{\bigg(1-\frac{8m_{Q}^{2}}{s}\bigg)^{2}-\frac{4m_{Q}^{2}}{s}}\Bigg], \nonumber\\ \alpha_{max}&=&\frac{1}{2}\Bigg[\bigg(1-\frac{8m_{Q}^{2}}{s}\bigg)+\sqrt{\bigg(1-\frac{8m_{Q}^{2}}{s}\bigg)^{2}-\frac{4m_{Q}^{2}}{s}}\Bigg], \nonumber\\
\beta_{min}&=&\frac{1}{2}\Bigg[\Big(1+2\alpha-\frac{3\alpha^{2}s}{\alpha s-m_{Q}^{2}}\Big)-\sqrt{\frac{[\alpha(1-\alpha)s-m_{Q}^{2}][\alpha(1-\alpha)s-(1+8\alpha)m_{Q}^{2}]}{(\alpha s-m_{Q}^{2})^{2}}}\Bigg],\nonumber\\
\beta_{max}&=&\frac{1}{2}\Bigg[\Big(1+2\alpha-\frac{3\alpha^{2}s}{\alpha s-m_{Q}^{2}}\Big)+\sqrt{\frac{[\alpha(1-\alpha)s-m_{Q}^{2}][\alpha(1-\alpha)s-(1+8\alpha)m_{Q}^{2}]}{(\alpha s-m_{Q}^{2})^{2}}}\Bigg],\nonumber\\
\gamma_{min}&=&\frac{1}{2}\Bigg[(1-\alpha-\beta)-\sqrt{\frac{(1-\alpha-\beta)\Big\{(1-\alpha-\beta)[\alpha\beta s-(\alpha+\beta)m_{Q}^{2}]-4\alpha\beta m_{Q}^{2}\Big\}}{\alpha\beta s-(\alpha+\beta)m_{Q}^{2}}}\Bigg],\nonumber
\end{eqnarray}
and
\begin{eqnarray}
\gamma_{max}&=&\frac{1}{2}\Bigg[(1-\alpha-\beta)+\sqrt{\frac{(1-\alpha-\beta)\Big\{(1-\alpha-\beta)[\alpha\beta s-(\alpha+\beta)m_{Q}^{2}]-4\alpha\beta m_{Q}^{2}\Big\}}{\alpha\beta s-(\alpha+\beta)m_{Q}^{2}}}\Bigg].\nonumber
\end{eqnarray}

\section{Numerical analysis and discussions}\label{sec3}
In this part,
the heavy $m_{Q}$ is taken as the running charm mass
$m_{c}=1.27\pm0.02~\mbox{GeV}$ \cite{PDG},
and other input parameters are
$\langle g^{2}G^{2}\rangle=0.88\pm0.25~\mbox{GeV}^{4}$ and $\langle
g^{3}G^{3}\rangle=0.58\pm0.18~\mbox{GeV}^{6}$ \cite{svzsum,overview3,Narison}.
Following the standard criterion
of sum rule analysis, one could
find proper work windows for the threshold parameter $\sqrt{s_{0}}$ and the Borel
parameter $M^{2}$.
The lower bound of $M^{2}$ could be gained in view of the OPE
convergence, and
the upper one is obtained from the pole dominance.
Meanwhile, the threshold
$\sqrt{s_{0}}$ is around
$0.4\sim0.6~\mbox{GeV}$ higher than the extracted
$M_{H}$ empirically, for which describes the beginning
of continuum state.

Taking the scalar-scalar case an example,
the inputs are kept at their central values at the start.
To find the lower bound of $M^{2}$, its OPE
convergence is shown by comparing the relative contributions of various
condensates from sum rule (\ref{sumrule}) for $\sqrt{s_{0}}=7.0~\mbox{GeV}$ in FIG. 1,
and one could note that the relative contributions of
two-gluon condensate $\langle
g^{2}G^{2}\rangle$ and three-gluon
condensate $\langle g^{3}G^{3}\rangle$
are very small.
Numerically, it
is taken as $M^{2}\geq2.5~\mbox{GeV}^{2}$
with an eye to the OPE convergence analysis.
On the other hand, the upper one of $M^{2}$ is obtained from
the pole contribution dominance phenomenologically.
FIG. 2 makes the comparison
between pole and continuum contribution from sum rule (\ref{sumrule})
for $\sqrt{s_{0}}=7.0~\mbox{GeV}$.
The relative pole contribution
is about $50\%$ at $M^{2}=3.0~\mbox{GeV}^{2}$ and decreasing with $M^{2}$.
Thereby, it could satisfy the pole dominance requirement
while $M^{2}\leq3.0~\mbox{GeV}^{2}$, and the Borel window is fixed as
$M^{2}=2.5\sim3.0~\mbox{GeV}^{2}$ for $\sqrt{s_0}=7.0~\mbox{GeV}$.
Analogously, they are taken as $M^{2}=2.5\sim2.9~\mbox{GeV}^{2}$
for $\sqrt{s_0}=6.9~\mbox{GeV}$, and
 $M^{2}=2.5\sim3.2~\mbox{GeV}^{2}$ for $\sqrt{s_0}=7.1~\mbox{GeV}$, respectively.
The mass $M_{H}$ dependence on $M^{2}$
is shown in FIG. 3 for the scalar-scalar case, and it is computed to be
$6.44\pm0.13~\mbox{GeV}$ in the chosen windows.
And then varying all the input values, the attained mass is
$6.44\pm0.13^{+0.02}_{-0.03}~\mbox{GeV}$ (the first error from
variation of $s_{0}$ and
$M^{2}$, and the second from the uncertainty of QCD parameters)
or $6.44^{+0.15}_{-0.16}~\mbox{GeV}$ in a short form.

In the very similar analyzing processes, proper work windows
for other three cases
could also be found and their corresponding Borel curves
are respectively given in FIG. 4--6.
After considering the uncertainty from both work windows
and variation of input parameters,
mass values of $0^{+}$ $cc\bar{c}\bar{c}$ tetraquark states are gained as
$6.45^{+0.14}_{-0.16}~\mbox{GeV}$ for the pseudoscalar-pseudoscalar configuration,
$6.46^{+0.13}_{-0.17}~\mbox{GeV}$ for the axial-axial one,
and  $6.47^{+0.12}_{-0.18}~\mbox{GeV}$ for the vector-vector one,
respectively.

\begin{figure}[htb!]
\centerline{\epsfysize=5.8truecm\epsfbox{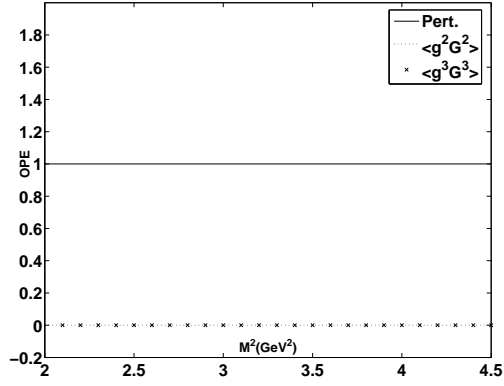}}
\caption{The OPE convergence for the $0^{+}$ fully-charmed tetraquark state
with a scalar-scalar configuration is shown by comparing the relative contributions of
perturbative, two-gluon condensate $\langle
g^{2}G^{2}\rangle$, and three-gluon
condensate $\langle g^{3}G^{3}\rangle$
from sum rule (\ref{sumrule})
for $\sqrt{s_{0}}=7.0~\mbox{GeV}$.}
\end{figure}

\begin{figure}
\centerline{\epsfysize=5.8truecm\epsfbox{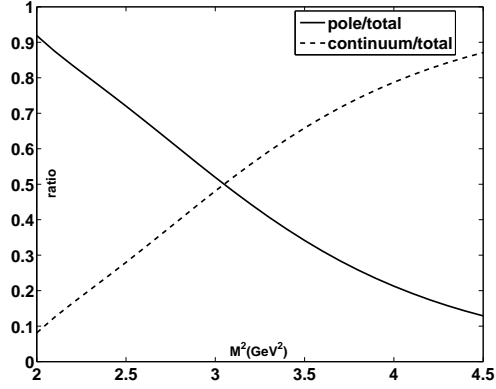}}
\caption{The phenomenological contribution in sum rule
(\ref{sumrule}) for $\sqrt{s_{0}}=7.0~\mbox{GeV}$ for
the $0^{+}$ fully-charmed tetraquark state
with a scalar-scalar configuration.
The solid line is the relative pole contribution
as a function of $M^2$ and the dashed line is the relative continuum
contribution.}
\end{figure}

\begin{figure}
\centerline{\epsfysize=5.8truecm
\epsfbox{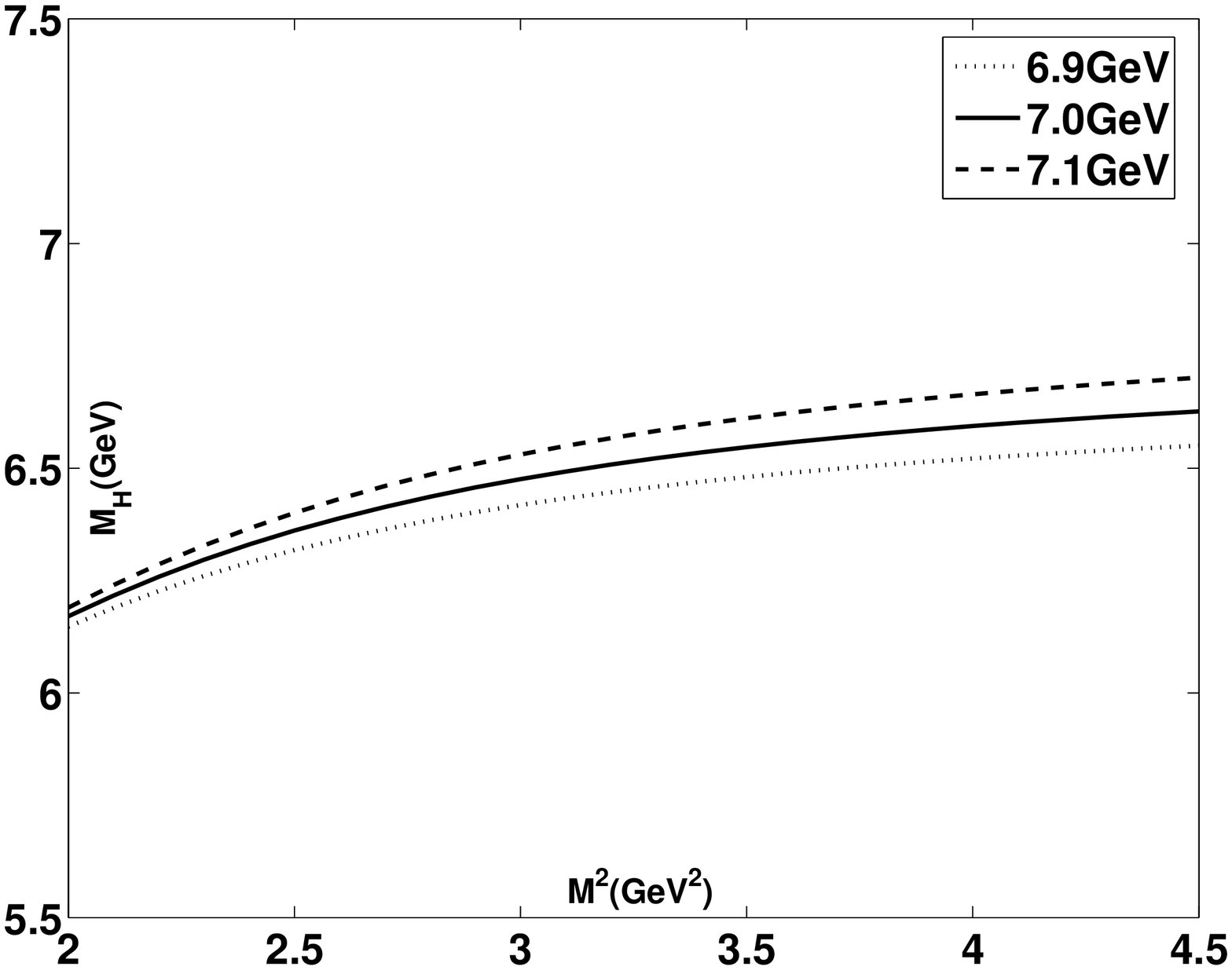}}\caption{The mass $M_{H}$
dependence on $M^2$ for the
$0^{+}$ fully-charmed tetraquark state
with a scalar-scalar configuration
 from sum rule (\ref{sum rule}) is shown.
The Borel windows of $M^{2}$ are $2.5\sim2.9~\mbox{GeV}^{2}$ for
$\sqrt{s_0}=6.9~\mbox{GeV}$, $2.5\sim3.0~\mbox{GeV}^{2}$
for $\sqrt{s_0}=7.0~\mbox{GeV}$, and
$2.5\sim3.2~\mbox{GeV}^{2}$ for $\sqrt{s_0}=7.1~\mbox{GeV}$, respectively.}
\end{figure}

\begin{figure}
\centerline{\epsfysize=5.8truecm
\epsfbox{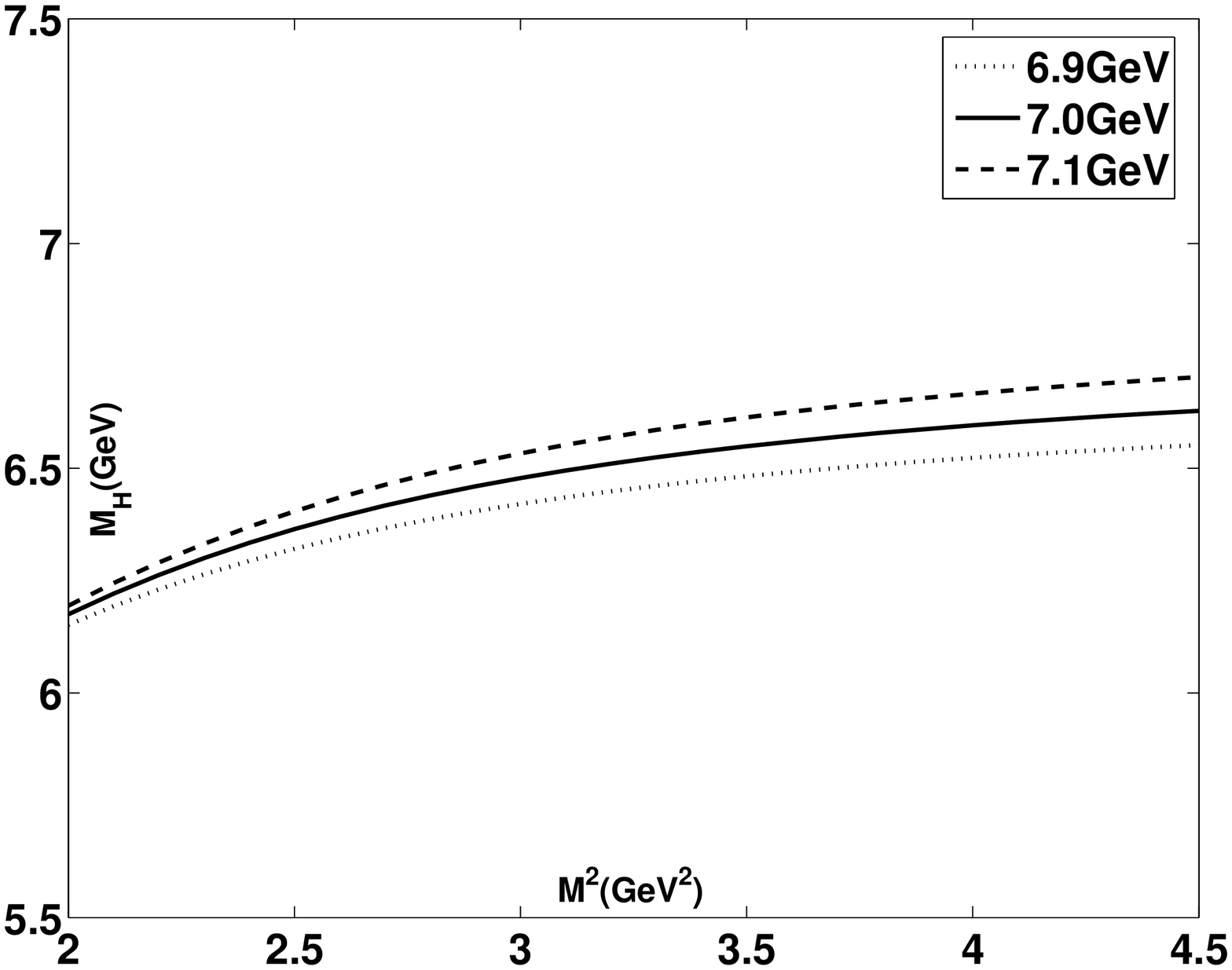}}\caption{The
mass $M_{H}$ dependence on $M^2$ for
the $0^{+}$ fully-charmed tetraquark state
with a pseudoscalar-pseudoscalar configuration
 from sum rule (\ref{sum rule}) is shown.
The Borel windows of $M^{2}$ are $2.5\sim2.9~\mbox{GeV}^{2}$ for
$\sqrt{s_0}=6.9~\mbox{GeV}$, $2.5\sim3.0~\mbox{GeV}^{2}$
for $\sqrt{s_0}=7.0~\mbox{GeV}$, and
$2.5\sim3.2~\mbox{GeV}^{2}$ for $\sqrt{s_0}=7.1~\mbox{GeV}$, respectively.}
\end{figure}

\begin{figure}
\centerline{\epsfysize=5.8truecm
\epsfbox{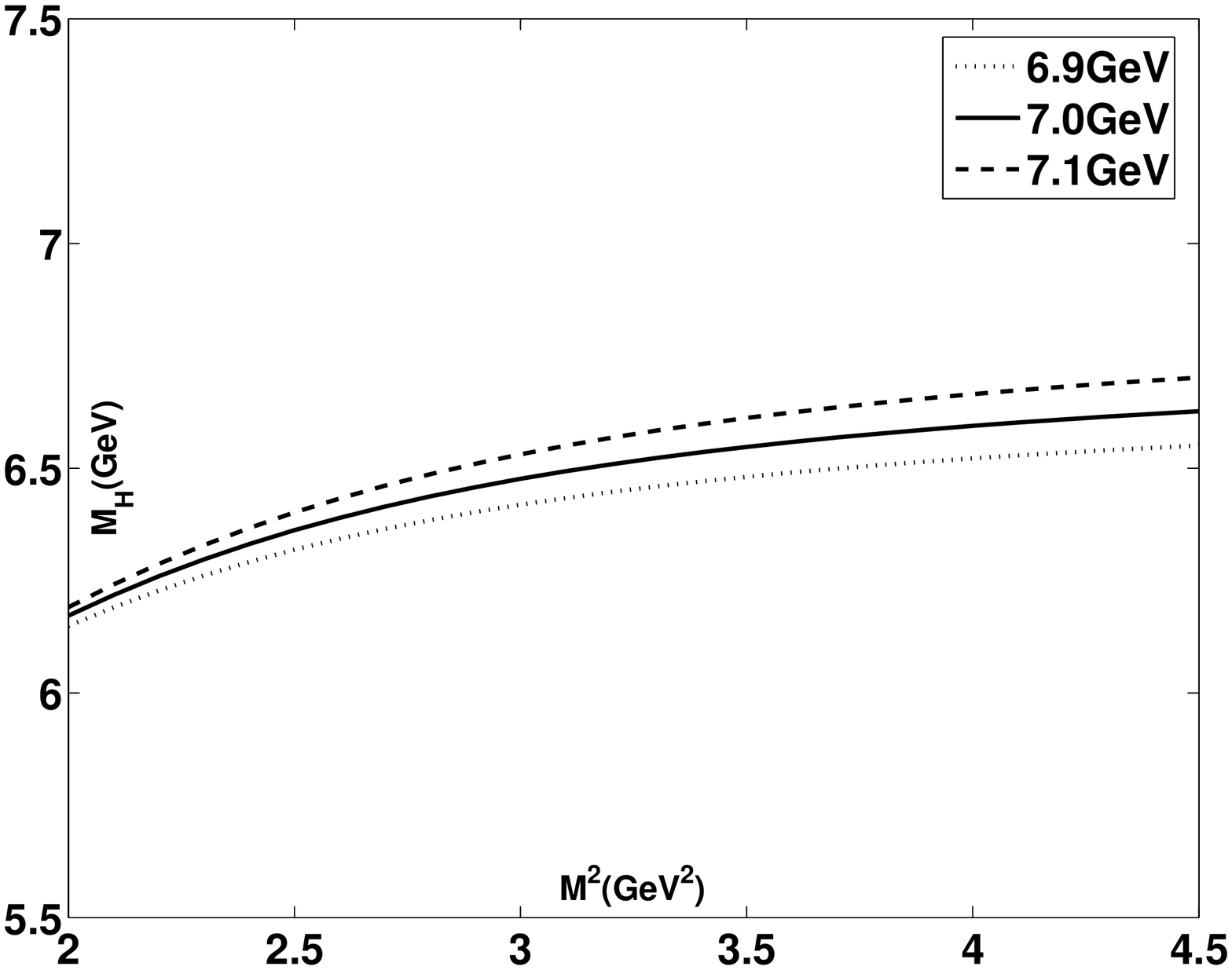}}\caption{The
mass $M_{H}$ dependence on $M^2$ for
the $0^{+}$ fully-charmed tetraquark state
with a axial-axial configuration
 from sum rule (\ref{sum rule}) is shown.
The Borel windows of $M^{2}$ are $2.5\sim2.9~\mbox{GeV}^{2}$ for
$\sqrt{s_0}=6.9~\mbox{GeV}$, $2.5\sim3.0~\mbox{GeV}^{2}$
for $\sqrt{s_0}=7.0~\mbox{GeV}$, and
$2.5\sim3.2~\mbox{GeV}^{2}$ for $\sqrt{s_0}=7.1~\mbox{GeV}$, respectively.}
\end{figure}

\begin{figure}
\centerline{\epsfysize=5.8truecm
\epsfbox{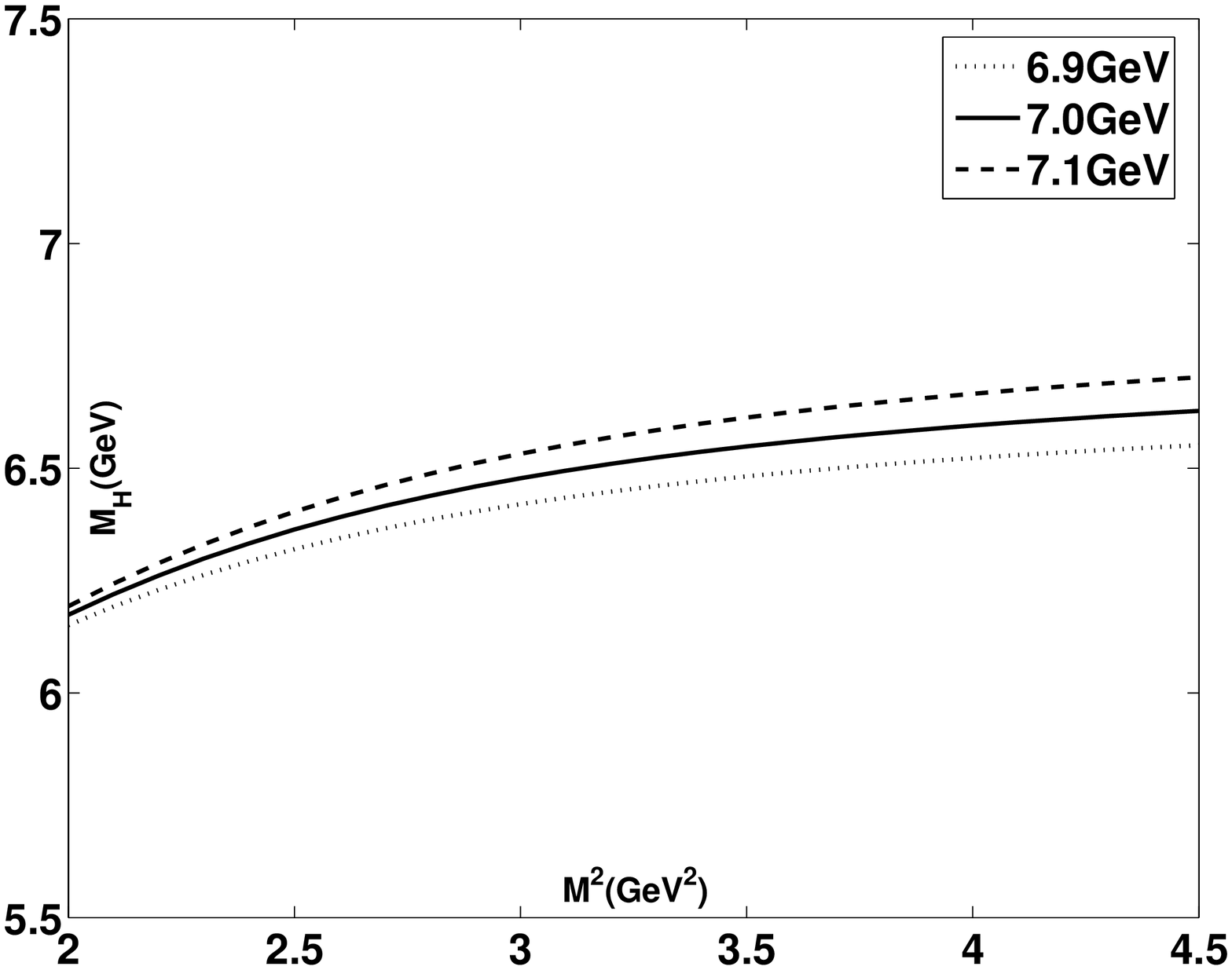}}\caption{The
mass $M_{H}$ dependence on $M^2$ for
the $0^{+}$ fully-charmed tetraquark state
with a vector-vector configuration
from sum rule (\ref{sum rule}) is shown.
The Borel windows of $M^{2}$ are $2.5\sim2.9~\mbox{GeV}^{2}$ for
$\sqrt{s_0}=6.9~\mbox{GeV}$, $2.5\sim3.0~\mbox{GeV}^{2}$
for $\sqrt{s_0}=7.0~\mbox{GeV}$, and
$2.5\sim3.2~\mbox{GeV}^{2}$ for $\sqrt{s_0}=7.1~\mbox{GeV}$, respectively.}
\end{figure}

\section{Summary}\label{sec4}
Focusing on the LHCb's new observation
in the di-$J/\psi$ mass spectrum,
we systematically investigate
$0^{+}$ fully-charmed tetraquark states in the framework of QCD sum rules.
By developing
related calculation techniques to fourfold heavy tetraquark states, four types of currents with different
configurations are taken into consideration and condensates up to dimension
$6$ are involved in the OPE side.
At last, mass values of $0^{+}$ $cc\bar{c}\bar{c}$ tetraquark states
 are calculated to be $6.44^{+0.15}_{-0.16}~\mbox{GeV}$
for the scalar-scalar case,
$6.45^{+0.14}_{-0.16}~\mbox{GeV}$ for the pseudoscalar-pseudoscalar case,
$6.46^{+0.13}_{-0.17}~\mbox{GeV}$ for the axial-axial case,
and  $6.47^{+0.12}_{-0.18}~\mbox{GeV}$ for the vector-vector case,
respectively.
All these results are numerically consistent with
the LHCb's experimental data
$6.2\sim6.8~\mbox{GeV}$ for the broad structure, which
could support its internal structure as a $0^{+}$
$cc\bar{c}\bar{c}$ tetraquark state.
For the future, it is expected that further
experimental and theoretical efforts
may reveal more on the nature of the exotic
states.

\begin{acknowledgments}
The author is very grateful to Xiang Liu
for first bringing the information
``Latest results on exotic hadrons at LHCb"
to his attention, and also thanks Zhi-Gang Wang for recent communication
and discussion.
This work was supported by the National
Natural Science Foundation of China under Contract
Nos. 11475258 and 11675263, and by the project for excellent youth talents in
NUDT.
\end{acknowledgments}


\end{document}